\documentclass[namedreferences]{solarphysics}
\usepackage[hyperref,optionalrh,solaromanenum]{spr-sola-addons} 
\usepackage{graphicx}                    
\usepackage{amssymb}                     
\usepackage{color}                       
\usepackage{breakurl}                    
                        
\usepackage{soul}

\newcommand{\degree}{\ensuremath{^\circ}}

\begin{document}
\begin{opening}
\title{Overlapping Magnetic Activity Cycles and the Sunspot Number: Forecasting Sunspot Cycle 25 Amplitude}

\author[addressref={1},corref,email={mscott@ucar.edu}]{\inits{S.W.}\fnm{Scott W.}~\lnm{McIntosh}~\orcid{0000-0002-7369-1776}}

\author[addressref={2}]{\inits{S.}\fnm{Sandra}~\lnm{Chapman}~\orcid{0000-0003-0053-1584}}

\author[addressref={3,4}]{\inits{R.J.}\fnm{Robert J.}~\lnm{Leamon}~\orcid{0000-0002-6811-5862}}

\author[addressref={1}]{\inits{R.E.}\fnm{Ricky}~\lnm{Egeland}~\orcid{0000-0002-4996-0753}}

\author[addressref={2,5,6}]{\inits{N.W.}\fnm{Nicholas W.}~\lnm{Watkins}~\orcid{0000-0003-4484-6588}}

\address[id={1}]{National Center for Atmospheric Research, High Altitude Observatory, P.O. Box 3000, Boulder, CO~80307, USA}
\address[id={2}]{Centre for Fusion, Space and Astrophysics, University of Warwick, Coventry CV4~7AL, UK}
\address[id={3}]{University of Maryland--Baltimore County, Goddard Planetary Heliophysics Institute, Baltimore, MD 21250, USA}
\address[id={4}]{NASA Goddard Space Flight Center, Code 672, Greenbelt, MD 20771, USA}
\address[id={5}]{Centre for the Analysis of Time Series, London School of Economics and Political Science, London WC2A 2AZ, UK}
\address[id={6}]{School of Engineering and Innovation, STEM Faculty, The Open University, Milton Keynes, UK}

\begin{abstract}
The Sun exhibits a well-observed modulation in the number of spots on its disk over a period of about 11 years. From the dawn of modern observational astronomy sunspots have presented a challenge to understanding -- their quasi-periodic variation in number, first noted 175 years ago, stimulates community-wide interest to this day. A large number of techniques are able to explain the temporal landmarks, (geometric) shape, and amplitude of sunspot ``cycles,'' however forecasting these features accurately in advance remains elusive. Recent observationally-motivated studies have illustrated a relationship between the Sun's 22-year (Hale) magnetic cycle and the production of the sunspot cycle landmarks and patterns, but not the amplitude of the sunspot cycle. Using (discrete) Hilbert transforms on more than 270 years of (monthly) sunspot numbers we robustly identify the so-called "termination" events that mark the end of the previous 11-yr sunspot cycle, the enhancement/acceleration of the present cycle, and the end of 22-yr magnetic activity cycles. Using these we extract a relationship between the temporal spacing of terminators and the magnitude of sunspot cycles. Given this relationship and our prediction of a terminator event in  2020, we deduce that Sunspot Cycle 25 could have a magnitude that rivals the top few since records began. This outcome would be in stark contrast to the community consensus estimate of sunspot cycle 25 magnitude.
\end{abstract}

\end{opening}

\section*{Introduction}
\begin{figure}[!ht]
    \centering
    \includegraphics[width=1.0\textwidth]{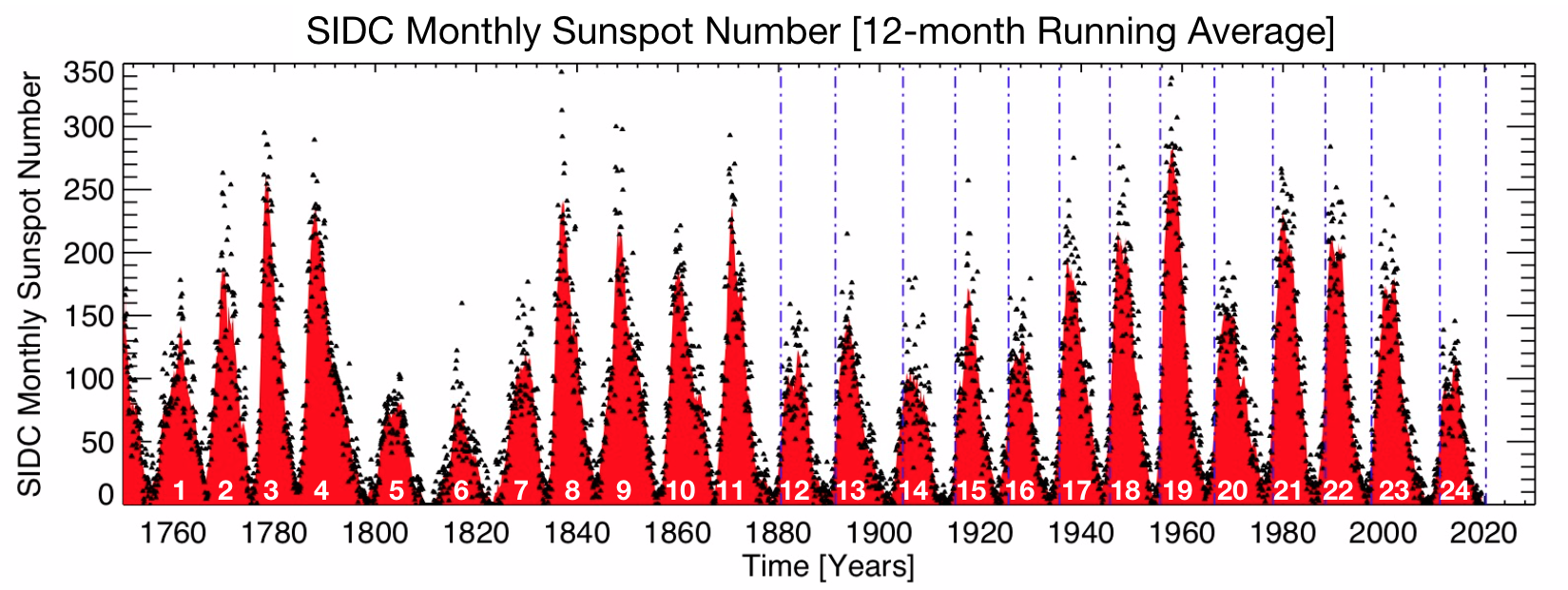}
    \caption{{\bf The monthly sunspot number since 1749.} The data values are represented by dots and the 12-month running average values are illustrated as a red shaded area. The sunspot cycle numbers are shown in the shaded area---number 1 starting in the 1755 and number 24 presently drawing to a close. Also shown in the figure are a set of vertical blue dashed lines that signify the magnetic activity cycle termination times that trigger the rapid growth of sunspot activity \citep{2019SoPh..294...88M}. The sunspot data used here are freely available and provided by the World Data Center-SILSO of the Royal Observatory of Belgium. We have used version 2.0 of the sunspot number \citep{2014SSRv..186...35C, 2015SpWea..13..529C} that is available at this website, identified as the ``Monthly mean total sunspot number'': \url{http://www.sidc.be/silso/}.}
    \label{f0}
\end{figure}

The (decadal) ebb and flow (waxing and waning) in the number of dark spots on the solar disk has motivated literally thousands of investigations since the discovery of the eponymous quasi-periodic 11-year sunspot cycle by \citep[][Fig.~\ref{f0}]{1844AN.....21..233S}. Since then, emphasis has been placed on determining the underlying physics of sunspot production \citep[e.g.,][]{2010LRSP....7....3C,2014ARA&A..52..251C,2015SSRv..196..101B,2017SSRv..210..367C} in addition to numerically forecasting the properties of upcoming cycles using statistical \citep[e.g.,][]{2018SpWea..16.1997P,2018SoPh..293..112P} or physical methods \citep[e.g.,][]{2018GeoRL..45.8091U,	
2018NatCo...9.5209B}. In recent decades, as the amplitude and timing of the sunspot cycle has reached greater societal significance, community-wide panels have been convened and charged with constructing consensus opinions on the upcoming sunspot cycle---several years in advance of the upcoming peak \citep{2008SoPh..252..209P}. Lack of adequate constraints, conflicting assumptions related to the solar dynamo mechanism, and different techniques, safe to say, result in a broad range of submissions to these panels that cover almost all potential ``physically reasonable'' outcomes \citep{2016SpWea..14...10P, 2020LRSP...17....2P}.

Sunspot cycle prediction is a high-stakes business and has become a decadal event, starting officially for Solar Cycle 23 \citep{1997EOSTr..78..205J}, and repeated for Solar Cycle 24 \citep{2008SoPh..252..209P}, the effort brought together a range of subject matter experts and an array of submitted methods that range from polar magnetic field precursors, through numerical models, and also using observed climatologies to extrapolate out in time \citep[e.g.,][]{ 2020LRSP...17....2P}. It is worth noting that the ``polar predictor'' method, which exploited measurements of the Sun's polar magnetic field at solar minimum as a proxy for the upcoming sunspot cycle strength, proved to be accurate for Solar Cycle 24 \citep[e.g.,][]{ 2005GeoRL..32.1104S,2005GeoRL..3221106S} and have informed much of the science that has followed.

Sunspot Cycle 25 is no different in terms of stakes \-- bringing some of the most sophisticated physical model forecasts to the discussion in addition to the robust and refined data-motivated methods \-- the international NOAA/NASA co-chaired Solar Cycle 25 Prediction Panel, (hereafter SC25PP) delivered the following consensus prognostication: Sunspot Cycle 25 (hereafter SC25) will be similar in size to Sunspot Cycle 24 (hereafter SC24). SC25 maximum will occur no earlier than the year 2023 and no later than 2026 with a minimum peak sunspot number\footnote{When quoting sunspot maxima we follow the convention of prediction panels past in this manuscript. Throughout, we quote the smoothed sunspot number for maxima, a value that is determined using a running thirteen month smoothing of the average number of sunspots for each calendar month.} of 95 and a maximum peak sunspot number of 130. Finally, the panel expects the end of SC24 and start of SC25 to occur no earlier than July, 2019, and no later than September, 2020\footnote{The interested reader can read the official NOAA press release describing the Panel's forecast at \url{https://www.weather.gov/news/190504-sun-activity-in-solar-cycle}. Although, we note that version 2.0 of the Sunspot number \citep{2015SpWea..13..529C}, and indicates that the peak smoothed sunspot number for cycle 24 was 116.}.

\begin{figure}[!ht]
    \centering
    \includegraphics[width=1.0\textwidth]{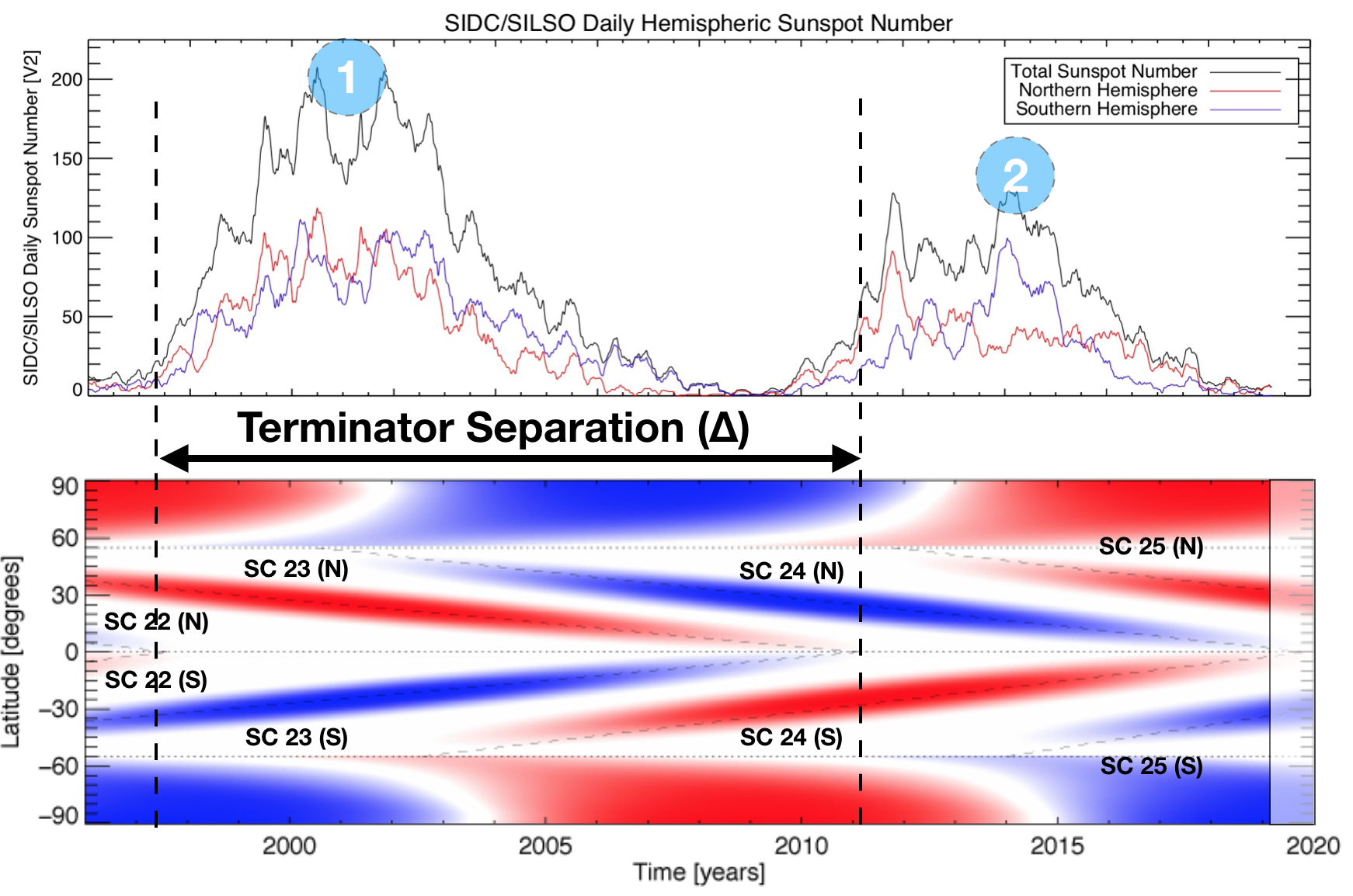}
    \caption{{\bf Inferred latitude versus time evolution of the magnetic activity bands and termination events of the 22-year Hale cycle over the past 22 years.} \emph{Top}: Hemispheric and total sunspot number of the recent cycles 23 and 24. Vertical lines show the termination events of cycle 22, 23, and (predicted) 24, which are followed by a rapid rise in solar activity. \emph{Bottom}: A conceptual drawing of the hypothesized activity bands of M2014 that are the underlying structure of the extended solar cycle. The indicated separation between the cycle 22 and 23 terminators provides a predictor for the cycle 24 amplitude, while likewise the separation between the observed cycle 23 terminator and the predicted cycle 24 terminator provides a method to forecast cycle 25. The numbered tags in the upper panel are illustrative for the reader and the two experiments that we will conduct below.}
    \label{fS1}
\end{figure}

\cite{2014ApJ...792...12M} (hereafter M2014) inferred that the sunspot cycle could be described in terms of the (magnetic) interactions of the oppositely polarized, spatio-temporally overlapping toroidal bands of the Sun's 22-year magnetic activity, or Hale, cycle (see, e.g., Fig.~\ref{fS1}). Those band interactions take place within a solar hemisphere and across the solar Equator. Further, they asserted that the degree by which the magnetic bands in the system temporally overlap defines the maximum amplitude of a sunspot cycle, the assumption being that there must be a sufficient amount of locally (or globally) imbalanced magnetic field to buoyantly form a sunspot. Therefore, epochs for which the time of band overlap is short would result in high amplitude cycles and conversely for epochs of longer band overlap. This is perhaps best illustrated in Fig.~\ref{fS1} and considering the nature of sunspot minima---four oppositely polarized bands are within 40\degree{} latitude of the equator, effectively nullifying the Sun's ability to form spots. 

The epoch immediately following sunspot cycle minimum conditions arises when the two lowest latitude bands cancel---the termination \citep[][hereafter M2019]{2019SoPh..294...88M}. In the picture of M2014, the termination of the old sunspot producing bands at the solar equator occurs at the end of their $\sim$19 year journey from $\sim$55\degree{} latitude and sees the Sun undergo a significant change in global magnetic activity on the scale of a single solar rotation. The termination signals the end of one sunspot (and magnetic) cycle and the start of the next sunspot cycle at mid-latitudes, acknowledging that the remaining bands of the magnetic cycle in each hemisphere have been present for several years before the termination (see, e.g., Fig.~\ref{fS1}) and the process which results in the reversal of the Sun's polar magnetic field.

M2014 explored only the last 60 years of solar activity, with only the later solar cycles including a high volume of EUV data, and so there was little attempt to quantify the relationship between band overlap, interaction, and the amplitude of sunspot cycles. In the picture of M2014, the temporal separation of the termination events can be used as a measure of band overlap. M2019 extended that analysis back another century such that 13 sunspot cycles had their terminator events identified \-- see the vertical blue dashed lines in Fig.~\ref{f1} and Table~\ref{T1}. 

Following M2019, \cite{2020SoPh..295...36L} explored an algorithmic approach to the identification of termination events in sunspot and activity proxy data. This was achieved by exploiting the properties of the discrete Hilbert transform \citep{1999ITSP...47.2604M}. They identified that the activity proxies displayed a common property in that the amplitude and phase functions that result from the discrete Hilbert transform peak and undergo a phase flip identically at the terminators---basically identifying the most rapid changes in the timeseries. They verified the termination events identified by M2014 and used their algorithmic approach to extend the terminator record back to 1820 (cycles 7--24), using the recently updated historical sunspot record \citep{2014SSRv..186...35C, 2015SpWea..13..529C}. Figure~\ref{f1} shows the reconstructed monthly sunspot number \citep{2015SpWea..13..529C} from which we will base the analysis presented here, exploiting the discrete Hilbert transform to explore the relationship between magnetic activity cycle band overlap (via terminator separation) and the amplitude of (resulting) sunspot cycles.

\subsection*{The Hilbert Transform} 
For a given time series $S(t)$ we can obtain an analytic signal \citep{Gabor1946} $A(t) exp[i \phi(t)]=S(t) +i H(t)$ where $H(t)$ is the Hilbert transform \citep{2000fta..book.....B} of $S(t)$ and  $A(t)$ and $\phi(t)$ are the analytic signal amplitude and phase respectively. For a discrete signal such as the monthly sunspot number analysed here, a discrete analytic signal can be constructed from the discrete Fourier transform of the original signal. We have used a standard method \citep{1999ITSP...47.2604M} which satisfies both invertability and orthogonality, as implemented in Matlab's {\tt hilbert} function. There are alternative ways to define instantaneous phases and amplitudes of the solar cycle considered by \cite{2002SoPh..208..167M}. however, they concluded that the analytic signal approach is best.

While defined for an arbitrary time series, the analytic signal will only give a physically meaningful decomposition of the original time series if the instantaneous frequency $\omega(t) =d\phi(t)/dt$  remains positive \citep{1992IEEEP..80..520B}. For a positive-definite signal such as the monthly sunspot number we therefore need to remove a background trend (see \citet{doi:10.1063/1.5025333} for an example, and further discussions in \cite{2002AmJPh..70..655P}, \cite{1992IEEEP..80..520B} and \cite{1998RSPSA.454..903H}). We obtained a slowly-varying trend by performing a robust local linear regression which down-weights outliers (``rlowess'') using Matlab's {\tt smooth} function with a 40 year window. 

\subsection*{Determination of Terminator Dates} 
\begin{figure}[!ht]
    \centering
    \includegraphics[width=1.0\textwidth]{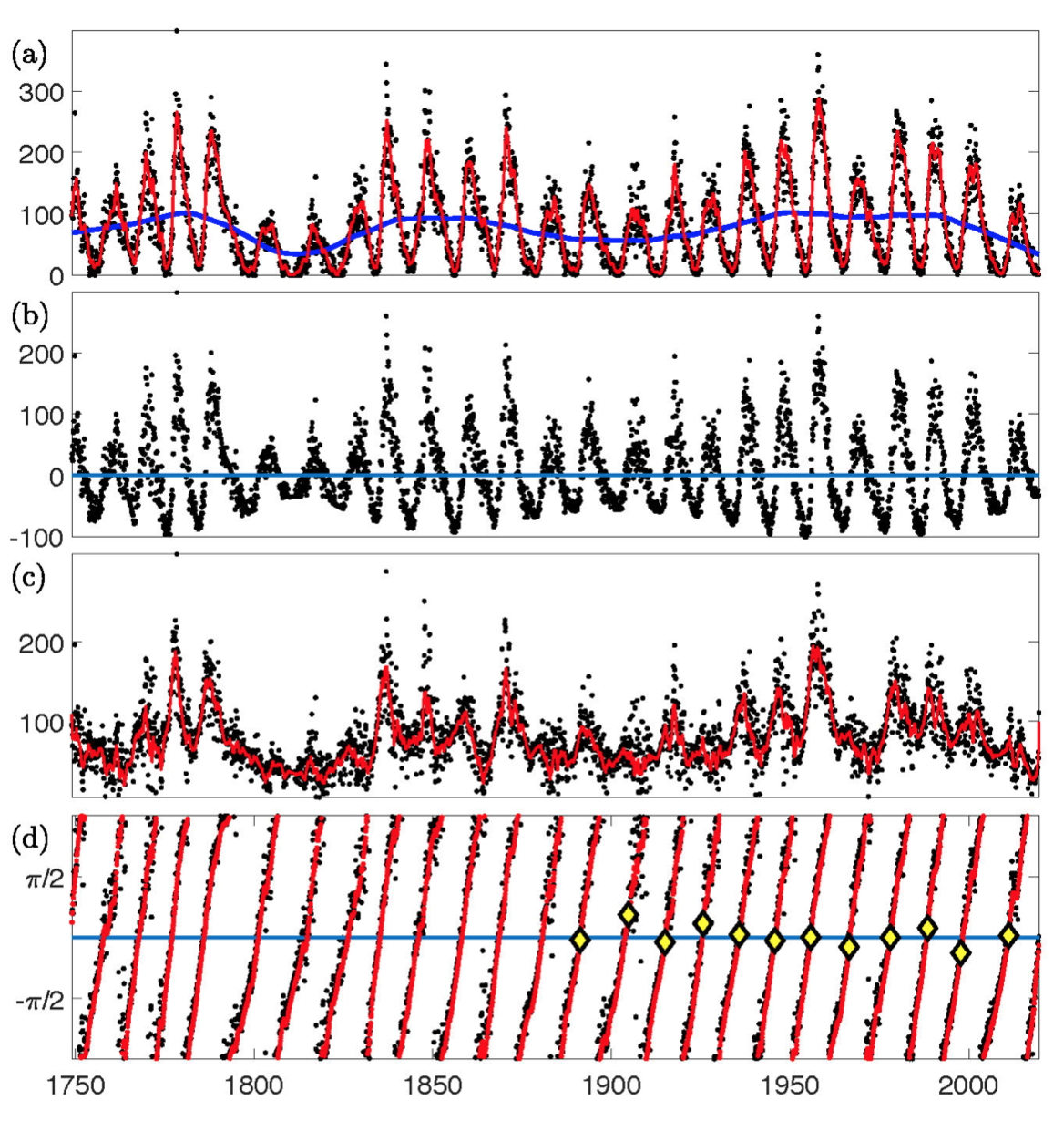}
    \caption{{\bf Discrete Hilbert transform of the monthly sunspot number since 1749.} From top to bottom (a) Monthly sunspot number (black), 12 month moving average (red) and slow timescale trend obtained by local regression using weighted linear least squares on a 40 year window (blue); (b) monthly sunspot number with local regression trend subtracted; (c) analytic signal amplitude of monthly (black) and 12 month moving average (red) sunspot number; (d) analytic signal phase as in (c), the yellow diamonds indicate terminators obtained previously\cite{2019SoPh..294...88M}. The terminator times used here are the analytic phase zero crossings.}
    \label{f1}
\end{figure}

We use the method discussed in \cite{2020GeoRL..4787795C} and  \cite{2020SoPh..295...36L} to determine our evaluation of the terminator dates via the analytic phase of the discrete Hilbert transform. For a given finite segment of a time series, the discrete Hilbert transform yields a difference in analytic phase relative to that at some (arbitrary) start time. For convenience here we have set zero phase to that at the terminator for the start of sunspot cycle 24 \citep{2019SoPh..294...88M}. In construction of Fig.~\ref{f1} we first performed a 12 month moving average of the monthly sunspot number and then computed the corresponding discrete analytic signal. The signal analytic phase was then linearly interpolated to obtain the phase zero crossings and the corresponding terminator times. The differences between successive terminator times do not depend on the choice of zero phase.

\section*{Results}
Performing a discrete Hilbert transform analysis and terminator identification \citep[see][]{2020SoPh..295...36L} but with the monthly (as opposed to daily) sunspot record and with the subtraction of a slowly time-varying trend as shown in panel~(a) of Fig.~\ref{f1}, permits the expansion of the terminator timeseries back to 1749. 
Indeed, such an analysis covers the ``Dalton'' minimum (from 1790 to 1830, or SC5 through SC7) in addition to the epochs of high activity in the late 1700s, 1850s and 1950s. In short, this period samples many of the solar activity extrema over the time of detailed human observation and cataloging.
Figure~\ref{f1}a shows the monthly sunspot number from 1749 until the present, as per Fig.~\ref{f0}. The red curve shows a 12 month boxcar smoothed version of the timeseries. The blue curve shown in panel~\ref{f1}a shows the local regression smoothing of the sunspot timeseries, where the smoothing window is chosen to be 40 years. Removing the smoothed sunspot trend from the monthly and 12-month smoothed timeseries results in the timeseries shown in panel~\ref{f1}b.
In \cite{2020SoPh..295...36L} we apply the discrete Hilbert transform to the two sunspot trend-subtracted timeseries to reveal the corresponding amplitude and phase functions of the discrete Hilbert transform in panels~\ref{f1}c and~\ref{f1}d, respectively. The application of the local regression smoothing to the timeseries results in a discrete Hilbert transforms that maintains a real-valued phase function. In contrast to the application of \cite{2020SoPh..295...36L} we have set the phase function of the discrete Hilbert transform to be identically zero at the terminator of 2011, meaning that zero crossings of the phase function, which are also coincident with maxima in the amplitude function, signify terminators in the timeseries. For reference, the terminators of M2014/M2019 are indicated as yellow diamonds. Notice the strong correspondence between the M2014/M2019 terminators and those derived independently here using the coarser (monthly) sunspot data.

Applying this methodology effectively doubles the number of terminators available for extended study. A visual comparison of Fig.~\ref{f1}d and~\ref{f1}a hint at a relationship between the separation of terminators and sunspot cycle amplitudes: low amplitude sunspot cycles appear to correspond with widely separated terminators while larger amplitude sunspot cycles correspond to more narrowly separated terminators. Table~\ref{T1} provides the sunspot cycle amplitudes, terminator dates and the length of the sunspot cycle derived from the separation of terminator events ($\Delta T$) that are derived from Figs.~\ref{f0} and~\ref{f1}.

\begin{figure}[!ht]
\centering
  \includegraphics[width=0.95\linewidth]{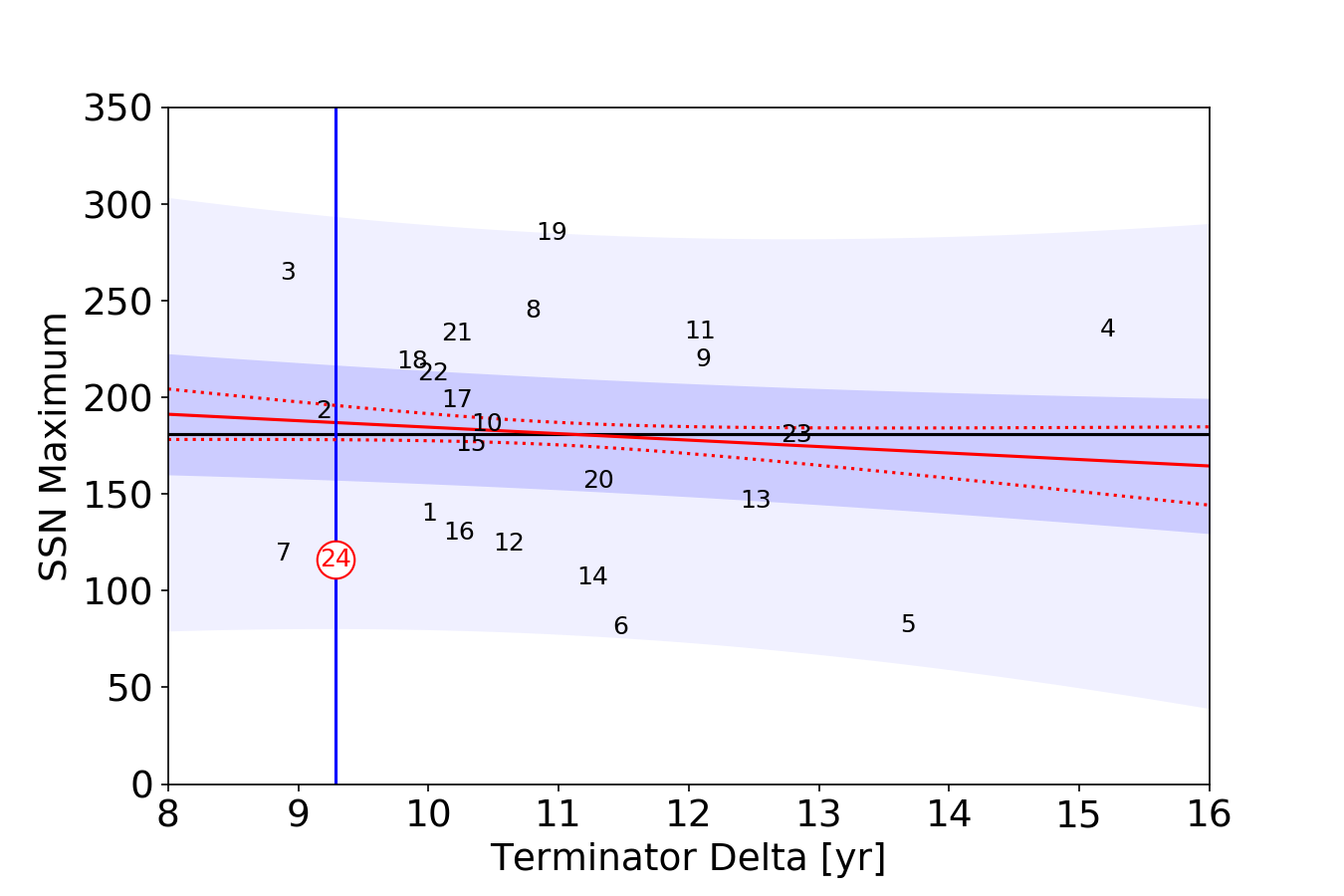}
  \includegraphics[width=0.95\linewidth]{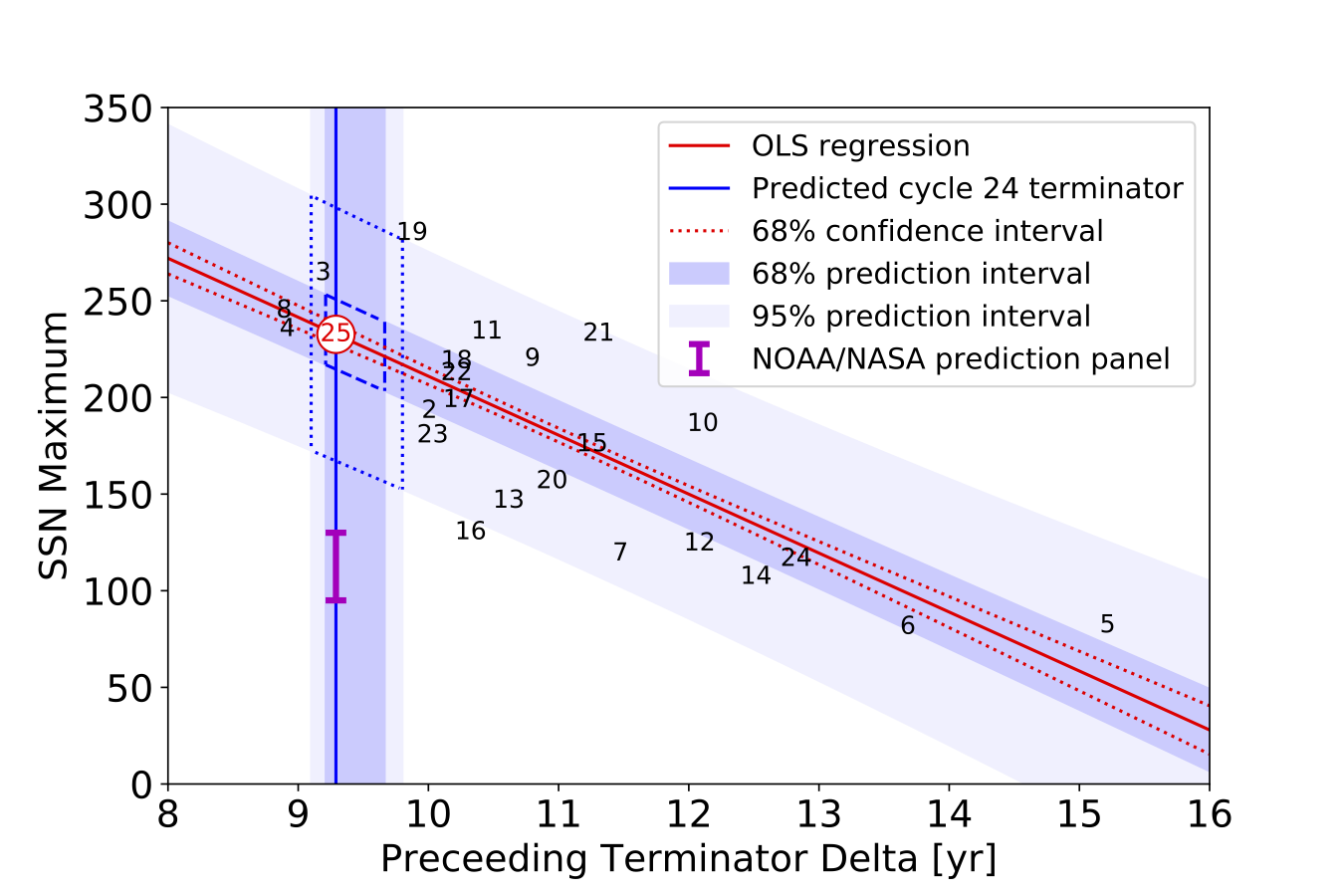}
\caption{{\bf Looking at relationships between terminator separation and sunspot cycle amplitudes.} Linear regressions of the terminator separation vs. (comparison 1; top panel) intermediate cycle sunspot maximum and (comparison 2; bottom panel) the following sunspot cycle maximum. The $1\sigma$ (68\%) confidence interval, as well as the $1\sigma$ (68\%) and $2\sigma$ (95\%) prediction intervals are shown. The predicted terminator separation for SC24 is shown in both panels, which along with the regression line results in a prediction for the amplitude of SC25 in panel~(b) that is significantly higher than the consensus prediction of the SC25PP (magenta bar). The black horizontal line in the top panel is the mean of SSN maximum, while the dashed and dotted blue lines in the bottom panel are the 68\% and 95\% prediction interval boundaries for the SC25 prediction, respectively.}
\label{f2}
\end{figure}

To explore this visual comparison, we analyzed the relationship between $\Delta T$ and the amplitude of that sunspot cycle and the upcoming ({\em i.e.}, next) sunspot cycle (see Fig.~\ref{fS1}). As demonstrated in the top panel of Fig.~\ref{f2}, we found no significant correlation between the terminator separation and the amplitude of the sunspot cycle it contains. The 68\% ($1\sigma$) confidence interval is shown to contain the zero-slope mean sunspot number (SSN) amplitude (black line), indicating the the null hypothesis of zero correlation is not rejected.

Compare now the terminator separation and the amplitude of the upcoming sunspot cycle that is shown in the bottom panel of Fig.~\ref{f2}. An ordinary least squares (OLS) regression, shows a significant anti-correlation between the two properties. 
The regression line is 
${\rm SSN}_{n+1} = (-30.5 \pm 3.8) \, \Delta T_n + 516$. 
The Pearson correlation coefficient is $r = -0.795$ and the correlation is significant to the 99.999\% level. We estimated the prediction intervals at 68\% ($1\sigma$) and 95\% ($2\sigma$) levels, which are also plotted in the bottom panel of Fig.~\ref{f2}. 

Using a SC24 terminator timing prediction of \cite{2020SoPh..295...36L} of $2020.37 ^{+0.38}_{-0.08} \, (1\sigma)$ along with our regression line and prediction intervals, our best estimate for the amplitude of SC25 is SSN=233, with a 68\% confidence that the amplitude will fall between SSN=204 and~254. Using the timing of \cite{2020SoPh..295...36L} this would result in a prediction (with 95\% confidence) that the SC25 amplitude will fall between SSN=153 and~305. At the time of acceptance the terminator of SC24 has not yet been reached and is lengthening the SC23 - SC24 terminator separation. Therefore the SC25 amplitudes mentioned should be considered as preliminary, and as limiting values decreasing by about 10 for every quarter year extension in the terminator separation. It is the intent of the authors to submit an clarification to the SC25 prediction presented herein when the SC24 terminator occurs. 

To put these values in perspective, and to highlight the strength of the relationship developed above, Fig.~\ref{f3} illustrates the SC25 forecast (at the 68\% confidence level) in purple, placed in contrast with that of the SC25PP consensus in green. The lightly shaded rectangle helps to place our forecast in contrast with past sunspot cycles---as projected SC25 would be in the top five of those observed. Further, the red dots in the plot are reconstructions, or a hindcast, of the solar maximum amplitudes given only the measured values of $\Delta T$ and the relationship derived above. With the exception of under-predicting the amplitude of SC 10, 19, and 21 (recall that the values used to develop the bottom panel of Fig.~\ref{f2} are drawn from annually smoothed data) the recovery of the sunspot maxima is very encouraging although we note that it appears to systematically {\em underestimate\/} the larger amplitude sunspot cycles.

\subsection*{Impact of Smoothing Windows on Terminator Dates}
In the development of the material above we have investigated how the two smoothing parameters (the timescale over which the trend is developed and the shorter timescale smoothing applied to the trend-removed residual time series; `trend' and `residual' respectively, for short) can influence  determination of the terminator. We also considered the impact of running-mean versus rlowess statistic in construction of the timeseries trend: for the parameter set used in Fig.~\ref{f1} the difference between the inferred terminators between these two approaches was 0.03\%.

Given this, and the application of the rlowess statistic in the above figures, we will use it in the estimation of the smoothing impact analysis. With the analysis of \cite{2020SoPh..295...36L} in mind we have varied the width of the window used in the trend (for a range from 5--150 years). Below 15 years there are extra (false) terminator crossings at zero phase. Below 35 years there are notable ripples on the trend. Longer than 125 years we see the Hilbert phase fails to monotonically increase with time in places such that it no longer resolves weaker sunspot cycles. This effect begins to influence the analysis significantly beyond 60 years, therefore we will use a 35--60 year span as the working range for trend removal where there is very little impact ($<$0.1\%) on the derived terminators and, hence, the relationships of Fig.~\ref{f2}.

Similarly, fixing the trend to 40 years, we have studied the impact on terminator determination by varying the residual smoothing from 1 to 84 months. For residual time series smoothing of longer than 3 months the terminators and their separation are stable, but for larger values ($>$42 months), the terminator and terminator separations are more variable. Beyond 42 months the residual is over-smoothed and stops resembling the original input time series. Longer than 84 months the method fails to resolve weak sunspot cycles. We conclude that the applicable working range for residual smoothing is 9--30 months.

The output of these experiments can be used to evaluate their impact on the terminator separations and hence on the relationship identified in the bottom panel of Fig.~\ref{f2}. The $\Delta T$ vs sunspot maxima relationship has a small variance ($<$2\%) and an even smaller effect on the projected magnitude of SC25 ($<$0.5\% at 1-$\sigma$, and $<$0.9\% at 2-$\sigma$).


\section*{Contrasting With Solar Minimum Correlation Studies}
While not an exhaustive survey of the literature, we briefly compare the terminator separation--cycle amplitude relationship shown above with prominent works in the literature that use the separation of solar minima as a measure of cycle length to develop predictability of upcoming cycle strength. We have strong reservations about the latter given the discussion above on the overlapping nature of Hale cycles, their impact on the sunspot cycle (M2014), and especially in the context of solar minimum conditions that result from the mutual cancellation of four magnetic bands---not to mention the subjectivity of picking {\it when\/} sunspot minimum occurs \citep[e.g.,][]{2020LRSP...17....2P}. 

It has been previously noted that the amplitude of a sunspot cycle is anti-correlated to the duration of the previous sunspot cycle, as measured by the time duration between solar/sunspot minimum \citep[e.g.,][]{1954PASP...66..241C, 1994SoPh..151..177H}. For reference Hathaway's approach yielded a Pearson correlation coefficient of $r=-0.68$, while the earlier solar-minimum focused work of Chernosky had a Pearson correlation coefficient of $r=-0.71$. However, we point to the introduction (and Fig.~\ref{fS1}) that the concept of solar minimum is a physically ill-defined quantity whose value depends on the activity record and the smoothing method used \citep{2015LRSP...12....4H, 2020LRSP...17....2P}.

\begin{figure}[!ht]
    \centering
    \includegraphics[width=1.0\textwidth]{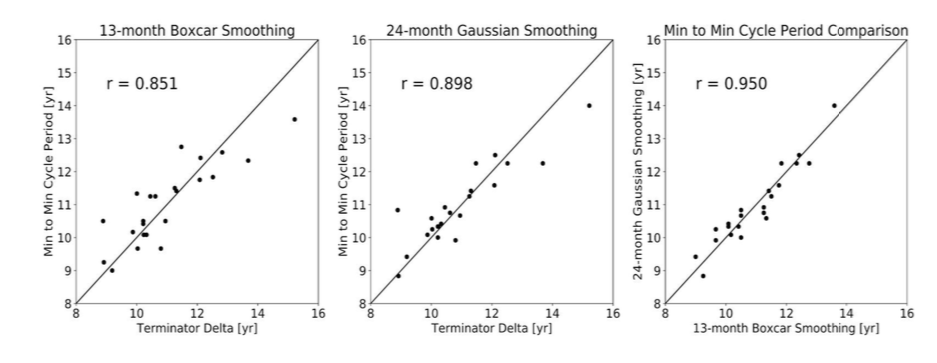}
    \caption{{\bf The relationship between terminator separation and sunspot minima separation.} Illustrating the relationships between sunspot minima separation versus terminator separation with different smoothing parameters used in the determination of the former, from left to right a 13-month boxcar smoothing method versus terminator separation, a 24-month Gaussian smoothing method versus terminator separation, and then the relationship between the two smoothing methodologies. In each case the Pearson correlation coefficient is shown.}
    \label{f4}
\end{figure}

\begin{figure}[!ht]
    \centering
    \includegraphics[width=1.0\textwidth]{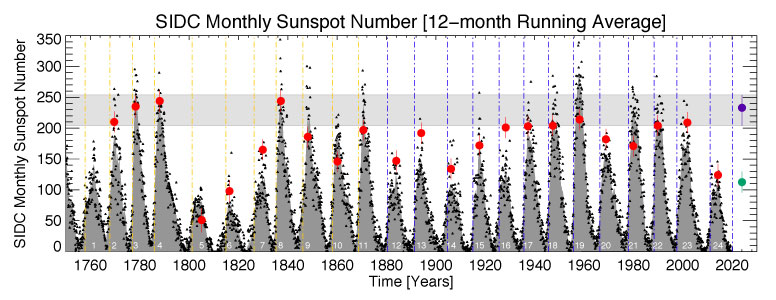}
    \caption{{\bf Sunspot Cycle 25 amplitude forecast in context.} The monthly mean sunspot number since 1749. The data values are represented by dots and the 12-month running average values are illustrated as a dark gray shaded area. The sunspot cycle numbers are shown in the shaded area---SC1 starting in the 1755 and SC24 presently drawing to a close. For comparison with Fig.~\ref{f0} we show the (M2019) vertical blue dashed lines that signify the magnetic cycle termination times that trigger the rapid growth of sunspot activity, while the vertical orange dashed lines show the discrete Hilbert transform derived terminators of \cite{2020SoPh..295...36L}, see Fig.~\ref{f1} and Tab.~\ref{T1}. Also shown are the forecast values of SC25 amplitude from the analysis above (purple dot) and the SC25PP (green dot). The horizontal light-gray shared region is to place the present forecast in historical context. Finally, we show the hindcast sunspot maxima for each cycle (red dots) derived from the measured terminator values and using the relationship developed above \-- the error bars on the hindcast dots represent the 68\% confidence value.}
    \label{f3}
\end{figure}

We contrast our approach taken above with a more traditional minimum\--to\--minimum cycle length versus terminator separations using two smoothing traditional cycle minimum determination methods: the 13-month boxcar smoothing (with half-weight endpoints) taken from Table~2 of \cite{2015LRSP...12....4H}; and a 24-month FWHM Gaussian smoothing from Table~1 of \cite{1999JGR...10422375H}. From left to right in Fig.~\ref{f4} we compare the 13-month boxcar smoothing with terminator separation, the 24-month boxcar FWHM Gaussian smoothing, and the inter-comparison of the two---along with their respective Pearson correlation coefficient. 

As one might expect, the terminator separation is well-correlated with cycle duration determined from the older methods. However, the two cycle minima methods are better correlated with each other than they are to the terminator separation. This indicates that a regression for minimum\--to\--minimum cycle length and amplitude of the upcoming cycle amplitude is not likely to improve for a specific choice of sunspot smoothing method, while the more robust terminator separation produces a better correlation than found in previous work.

For the interested reader, we identified several historical references while researching this section of the paper. \cite{2002SoPh..211..357H} points out \cite{1954PASP...66..241C} as previous work on the sunspot cycle amplitude-duration relationship. \cite{1954PASP...66..241C} in turn, pointed back to \cite{1861MNRAS..21...77W} as a previous source on the relationship for ``concurrent cycles'' ({\em i.e.}, as in Fig~\ref{f2}a), claiming it was ``one of the most important discoveries regarding solar conditions \citep[c.f.][]{1931ZA......2..370L}''. In this paper we use \cite{1954PASP...66..241C} as a motivator for the idea of a strong amplitude-period effect for ``following'' sunspot cycles ({\em i.e.}, Fig.~\ref{f2}b).

Finally, we compare the published Pearson correlation coefficients of the work published in these previous papers with the relationship derived herein:
\begin{itemize}
    \item{This work, the terminator separation method for SC1 through SC23 produces an {\it r-value} of $-0.795$}
    \item{Figure 29 of \cite{2015LRSP...12....4H}, using a 13-month boxcar smoothing for SC1 through SC23 produces an {\it r-value} of $-0.68$}
    \item{Figure 7a of \cite{2002SoPh..211..357H}, using a 24-month Gaussian smoothing for SC2 through SC22 produces an {\it r-value} of $-0.69$}
    \item{Figure 6 of \cite{1994SoPh..151..177H}, using sunspot cycle shape fit parameters for SC 2 through SC18) produces an {\it r-value} of $-0.63$}
    \item{Figure 1B of \cite{1954PASP...66..241C}, using 12-month mean smoothing for SC5 to SC18 produces an {\it r-value} of $-0.71$.}
\end{itemize}

Therefore, from this limited survey of the prominent literature on the topic, we see that terminator separation provides a statistically stronger indicator of sunspot cycle amplitude, being notably better than the solar minimum derived methods.

\section*{Discussion: Our Outside-The-Consensus Forecast}
The phenomenological model presented in M2014, and employed above, differs in one critical regard from the conventional physics-based models employed in the SC25PP, similar recently published efforts \citep{2018NatCo...9.5209B}, and now for machine-learning-inspired models \citep{2020ApJ...890...36K}. The common core feature of these models is that the magnetic fields present in, or generated by, them are dynamically passive with respect to the large-scale flows present in the system \citep{2010LRSP....7....3C}, or are ``frozen-in,'' using  magnetohydrodynamical terminology \citep{1942Natur.150..405A}. Conversely, an explanation for the hemispherically synchronized, rapid triggering of mid- and high-latitude magnetic flux emergence following termination events at the solar equator, requires that the magnetic bands of the Hale magnetic cycle are strong and are dynamically important relative to the flows \citep{2019SoPh..294...88M, 2019NatSR...9.2035D}. Finally, should there be strong divergence between the forecast presented above and those that utilize the polar predictor methodology \citep[e.g.,][]{ 2005GeoRL..32.1104S,2005GeoRL..3221106S} we should revisit the role of the Sun's polar magnetic field in the development of the Sun's dynamo mechanism.

Over the coming months, as SC25 matures, it will become evident which of these (very different) paradigms is most relevant - such is the contrast in the forecasts discussed herein. Very early indications of the spot pattern are appearing at higher than average latitudes \citep[$\sim$40\degree;][]{2020RNAAS...4...30N}. Historically, high latitude spot emergence has been associated with the development of large amplitude sunspot cycles \citep[e.g.,][]{1935MiZur..14..105W, 1939MiZur..14..439W, 2015LRSP...12....4H}---only time will tell how accurate all these predictions are for SC25.

\section*{Conclusion}
Our method predicts that SC25 could be among the strongest sunspot cycles ever observed, and that it will almost certainly be stronger than present SC24 (sunspot number of 116) and most likely stronger than the previous SC23 (sunspot number of 180). This is in stark contrast to the consensus of the SC25PP, sunspot number maximum between 95 and 130, {\em i.e.}, similar to that of SC24. Indeed, as can be seen in Fig.~\ref{f2}b, if our prediction for the 2020 terminator time is correct, such a low value would be a severe outlier with respect to the observed behavior of previous sunspot cycles. Such a low value could only be reconciled with the previously observed sunspot cycles if the next terminator event is delayed by more than two years from our predicted value, which would extend the present low activity levels to an extraordinary length. We note also that the relationship developed herein would have \emph{correctly\/} predicted the low amplitude of SC24 (from a terminators separation of 12.825 years) following the 2011 terminator---three years after the 2006 NOAA/NASA Solar Cycle Prediction Panel delivered their consensus prediction \citep{2008SoPh..252..209P}. Finally, the arrival of the SC24 terminator will permit higher fidelity on the forecast presented.

\subsection*{Data Availability}
The sunspot data used here are freely available and provided by the World Data Center-SILSO of the Royal Observatory of Belgium. We have used version 2.0 of the sunspot number \citep{2014SSRv..186...35C, 2015SpWea..13..529C} that is available at this website, identified as the ``Monthly mean total sunspot number'': \url{http://www.sidc.be/silso/}

\subsection*{Acknowledgments}
This material is based upon work supported by the National Center for Atmospheric Research, which is a major facility sponsored by the National Science Foundation under Cooperative Agreement No. 1852977. We thank Prof. Dibyendu Nandy for a critical reading of the paper and providing very useful feedback. SCC, NWW and RJL appreciate the support of the HAO Visitor Program. RJL acknowledges support from NASA's Living With a Star Program. SCC acknowledges AFOSR grant FA9550-17-1-0054. 

\subsection*{Author Contributions}
SMC devised and directed the experiment, was the primary author of the article supported by RJL, SCC, NWW, and RE. SCC and NWW proposed and performed the Hilbert Transform analysis presented. RJL and SCC devised and RE performed the statistical analysis of terminator separations and sunspot cycle amplitudes

\begin{table}[!h]
\caption{List of sunspot cycles, with their peak SSN, the times of their Terminators (rounded to the nearest month and expressed as a decimal year), and the difference  between subsequent terminators ($\Delta$T; yr). Recall that by definition the terminator of sunspot cycle $N$ occurs during the rise phase of sunspot cycle $N+1$.}
\label{T1}
 \begin{tabular}{cccr}
    \hline \hline
	SC &  Maximum & Terminator & $\Delta$T \\
	{} & Sunspot Number & Date & {} \\
	\hline 
       0 & \ldots &    1757.92 &       \ldots \\
       1 & 144 &       1767.92 &       10.00 \\
       2 & 193 &       1777.08 &        9.16 \\
       3 & 264 &       1786.00 &        8.92 \\
       4 & 235 &       1801.25 &       15.25 \\
       5 & 082 &       1814.92 &       13.66 \\
       6 & 081 &       1826.42 &       11.50 \\
       7 & 119 &       1835.25 &        8.83 \\
       8 & 245 &       1846.08 &       10.83 \\
       9 & 220 &       1858.17 &       12.08 \\
      10 & 186 &       1868.67 &       10.50 \\
      11 & 234 &       1880.75 &       12.08 \\
      12 & 124 &       1891.33 &       10.58 \\
      13 & 147 &       1903.83 &       12.50 \\
      14 & 107 &       1915.08 &       11.25 \\
      15 & 176 &       1925.42 &       10.33 \\
      16 & 130 &       1935.67 &       10.25 \\
      17 & 199 &       1945.92 &       10.25 \\
      18 & 219 &       1955.75 &        9.83 \\
      19 & 285 &       1966.67 &       10.92 \\
      20 & 157 &       1978.00 &       11.33 \\
      21 & 233 &       1988.25 &       10.25 \\
      22 & 213 &       1998.25 &       10.00 \\
      23 & 180 &       2011.08 &       12.83 \\
      24 & 116 &       \ldots &      \ldots \\ 
     \hline
  \end{tabular}
\end{table}

\bibliographystyle{spr-mp-sola}

\end{document}